\documentclass[preprint,aps]{revtex4}
\usepackage{amssymb,epsf}

\begin{document}

\title{Taub-NUT/Bolt Black Holes in Gauss-Bonnet-Maxwell Gravity}
\author{M. H. Dehghani$^{1,2}$\footnote{email address: mhd@shirazu.ac.ir} and S. H. Hendi$^{1}$}
\affiliation{$^1$Physics Department and Biruni Observatory,
College of Sciences, Shiraz University, Shiraz 71454, Iran\\
$^2$Research Institute for Astrophysics and Astronomy of Maragha
(RIAAM), Maragha, Iran}

\begin{abstract}
We present a class of higher dimensional solutions to
Gauss-Bonnet-Maxwell equations in $2k+2$ dimensions with a $U(1)$
fibration over a $2k$-dimensional base space $\mathcal{B}$. These
solutions depend on two extra parameters, other than the mass and
the NUT charge, which are the electric charge $q$ and the electric
potential at infinity $V$. We find that the form of metric is
sensitive to geometry of the base space, while the form of
electromagnetic field is independent of $\mathcal{B}$. We
investigate the existence of Taub-NUT/bolt solutions and find that
in addition to the two conditions of uncharged NUT solutions,
there exist two other conditions. These two extra conditions come
from the regularity of vector potential at $r=N$ and the fact that
the horizon at $r=N$ should be the outer horizon of the black
hole. We find that for all non-extremal NUT solutions of Einstein
gravity having no curvature singularity at $r=N$, there exist NUT
solutions in Gauss-Bonnet-Maxwell gravity. Indeed, we have
non-extreme NUT solutions in $2+2k$ dimensions only when the
$2k$-dimensional base space is chosen to be $\mathbb{CP}^{2k}$. We
also find that the Gauss-Bonnet-Maxwell gravity has extremal NUT
solutions whenever the base space is a product of 2-torii with at
most a $2$-dimensional factor space of positive curvature, even
though there a curvature singularity exists at $r=N$. We also find
that one can have bolt solutions in Gauss-Bonnet-Maxwell gravity
with any base space. The only case for which one does not have
black hole solutions is in the absence of a cosmological term with
zero curvature base space.
\end{abstract}

\maketitle

\section{Introduction}

The prominence of string theory as a theory of everything, in particular a
quantum theory of gravity, means that we should examine its consequences in
regimes where it departs from the Einstein gravity. One way of examining the
consequence of string theory on the solutions of classical gravity is
through the use of the field equations which arise from the effective action
of a low-energy limit of string theory. This effective action which
describes gravity at the classical level consists of the Einstein-Hilbert
action plus curvature-squared terms and higher powers as well, and in
general give rise to fourth order field equations and bring in ghosts \cite
{Wit1}. However, if the effective action contains the higher powers of
curvature in particular combinations, then only second order field equations
are produced and consequently no ghosts arise \cite{Zum}. The effective
action obtained by this argument is precisely of the form proposed by
Lovelock \cite{Lov}. It is therefore natural to suppose that the
construction of the Taub-Nut solutions of Gauss-Bonnet gravity, which is the
first order corrections of the string theory at low energy, might provide us
with a window on some interesting new corners of this theory. The first
attempt has been done by one of us, and the Taub-NUT/bolt solutions of
Gauss-Bonnet gravity have been constructed \cite{Deh1}. These solutions have
some features which are different from the Nut solutions of Einstein
gravity. Here, we construct the Taub-NUT solutions in Gauss-Bonnet-Maxwell
gravity and investigate their properties.

In the last decades a renewed interest appears in Lovelock
gravity. In particular, exact static spherically symmetric black
hole solutions of the Gauss-Bonnet gravity have been found in Ref.
\cite{Des}, and of the Maxwell-Gauss-Bonnet and
Born-Infeld-Gauss-Bonnet models in Ref. \cite{Wil1}. The
thermodynamics of the uncharged static spherically black hole
solutions has been considered in \cite{MS}, of solutions with
nontrivial topology in \cite{Cai} and of charged solutions in
\cite{Wil1,Od1}. All of these known solutions in Gauss-Bonnet
gravity are static. Recently one of us has introduced two new
classes of rotating solutions of second order Lovelock gravity and
investigate their thermodynamics \cite{Deh2}. Also, the exact
solutions in third order Lovelock gravity with the quartic terms
has been constructed recently \cite{Deh3}.

The original four-dimensional solution \cite{TNUT} is only locally
asymptotic flat. The spacetime has as a boundary at infinity a twisted $%
S^{1} $ bundle over $S^{2}$, instead of simply being $S^{1}\times S^{2}$. In
general, the Killing vector that corresponds to the coordinate that
parameterizes the fibre $S^{1}$ can have a zero-dimensional fixed point set
(called a NUT solution) or a two-dimensional fixed point set (referred to as
a `bolt' solution). There are known extensions of the Taub-NUT/bolt
solutions to the case when a cosmological constant is present. In this case
the asymptotic structure is only locally de Sitter (for positive
cosmological constant) or anti-de Sitter (for negative cosmological
constant) and the solutions are referred to as Taub-NUT-(A)dS metrics.
Generalizations to higher dimensions follow closely the four-dimensional
case \cite{Bais,Page,Akbar,Robinson,Awad,Lorenzo,Mann1,
Mann2,Astefan1,Astefan2}. Also, charged Taub-NUT solution of the
Einstein-Maxwell equations in four dimensions is known \cite{Bril}, and its
generalization to six dimensions has been done in Refs. \cite{Mann3,Awad2}.
The existence of NUT charged solutions of Einstein-Yang-Mills and
Einstein-Yang-Mills-Higgs theory and their thermodynamics have also been
considered \cite{Radu}. Dyonic Taub-NUT solution in the low energy limit of
string theory has also been investigated \cite{Myers1}.

In this paper we consider Taub-NUT solutions in Gauss-Bonnet-Maxwell gravity
in $2k+2$ dimensions. We find that NUT black holes exist, but
Gauss-Bonnet-Maxwell gravity introduces some features not present in
higher-dimensional Einstein-Maxwell gravity or Gauss-Bonnet gravity in the
absence of electromagnetic field. The form of the metric function is
sensitive to the base space over which the circle is fibred, while the form
of the electromagnetic field is independent of the base space. We find that
there exist some restrictions on the value of electric charge in order to
have NUT solutions. Furthermore, we confirm the two conjectures of Ref. \cite
{Deh1} and show that these conjectures can be extended to the case of
Gauss-Bonnet gravity in the presence of electromagnetic field. Indeed, we
show that pure non-extreme NUT solutions only exist if the base space has a
single factor of maximal dimensionality, and extreme NUT solutions exist if
the base space has at most one 2-dimensional curved space with positive
curvature as one of its factor spaces.

The outline of our paper is as follows. We give a brief review of the field
equations of second order Lovelock gravity in the presence of
electromagnetic field in Sec. \ref{Fiel}. In Sec. \ref{6d}, we obtain all
possible Taub-NUT/bolt solutions of Gauss-Bonnet-Maxwell gravity in six
dimensions. Then, in Secs. \ref{8d} and \ref{10d}, we present all kind of
Taub-NUT/bolt solutions of Gauss-Bonnet-Maxwell gravity in eight and ten
dimensions. In Sec \ref{dd}, we extend our study to the $(2k+2)$-dimensional
case. We finish our paper with some concluding remarks.

\section{Field Equations\label{Fiel}}

The most natural extension of general relativity in higher
dimensional spacetimes with the assumption of Einstein -- that the
left hand side of the field equations is the most general
symmetric conserved tensor containing no more than second
derivatives of the metric -- is Lovelock theory. The gravitational
action of this theory can be written as \cite{Lov}
\begin{equation}
I_{G}=\int d^{d}x\sqrt{-g}\sum_{n=0}^{[d/2]}\alpha _{k}{\cal L}_{k}
\label{Lov1}
\end{equation}
where $[z]$ denotes the integer part of $z$, $\alpha _{k}$ is an arbitrary
constant and ${\cal L}_{k}$ is the Euler density of a $2k$-dimensional
manifold,
\begin{equation}
{\cal L}_{k}=\frac{1}{2^{k}}\delta _{\rho _{1}\sigma _{1}\cdots \rho
_{k}\sigma _{k}}^{\mu _{1}\nu _{1}\cdots \mu _{k}\nu _{k}}R_{\mu _{1}\nu
_{1}}^{\phantom{\mu_1\nu_1}{\rho_1\sigma_1}}\cdots R_{\mu _{k}\nu _{k}}^{%
\phantom{\mu_k \nu_k}{\rho_k \sigma_k}}  \label{Lov2}
\end{equation}
In Eq. (\ref{Lov2}) $\delta _{\rho _{1}\sigma _{1}\cdots \rho _{k}\sigma
_{k}}^{\mu _{1}\nu _{1}\cdots \mu _{k}\nu _{k}}$ is the generalized totally
anti-symmetric Kronecker delta and $R_{\mu \nu }^{\phantom{\mu\nu}{\rho%
\sigma}}$ is the Riemann tensor. We note that in $d$ dimensions, all terms
for which $n>[d/2]$ are identically equal to zero, and the term $n=d/2$ is a
topological term. Consequently only terms for which $n<d/2$ contribute to
the field equations. Here we study Gauss-Bonnet gravity, that is first three
terms of Lovelock gravity. In this case the action is

\begin{equation}
I_{G}=\frac{1}{2}\int_{{\cal M}}d^{d}x\sqrt{-g}[-2\Lambda +R+\alpha (R_{\mu
\nu \gamma \delta }R^{\mu \nu \gamma \delta }-4R_{\mu \nu }R^{\mu \nu
}+R^{2})-F_{\mu \nu }F^{\mu \nu }]  \label{Act1}
\end{equation}
where $\Lambda $ is the cosmological constant, $\alpha $ is the Gauss-Bonnet
coefficient with dimension $({\rm length})^{2}$, $R$ and $R_{\mu \nu }$ are
the Ricci scalar and Ricci tensors of the spacetime, $F_{\mu \nu }=\partial
_{\mu }A_{\nu }-\partial _{\nu }A_{\mu }$ is electromagnetic tensor field
and $A_{\mu }$ is the vector potential. Since $\alpha $ is positive in
heterotic string theory \cite{Des} we shall restrict ourselves to the case $%
\alpha >0$. The first term is the cosmological term, the second term is just
the Einstein term, and the third term is the second order Lovelock
(Gauss-Bonnet) term. From a geometric point of view the combination of these
terms in five and six dimensions is the most general Lagrangian that yields
second order field equations, as in the four-dimensional case for which the
Einstein-Hilbert action is the most general Lagrangian producing second
order field equations.

Varying the action with respect to the metric tensor $g_{\mu \nu }$ and
electromagnetic tensor field $F_{\mu \nu }$ the equations of gravitation and
electromagnetic fields are obtained as:
\begin{eqnarray}
&&G_{\mu \nu }+\Lambda g_{\mu \nu }-\alpha \{4R^{\rho \sigma }R_{\mu \rho
\nu \sigma }-2R_{\mu }^{\ \rho \sigma \lambda }R_{\nu \rho \sigma \lambda
}-2RR_{\mu \nu }+4R_{\mu \lambda }R_{\text{ \ }\nu }^{\lambda }  \nonumber \\
&&+\frac{1}{2}g_{\mu \nu }(R_{\kappa \lambda \rho \sigma }R^{\kappa \lambda
\rho \sigma }-4R_{\rho \sigma }R^{\rho \sigma }+R^{2})\}=T_{\mu \nu }
\label{Geq}
\end{eqnarray}
\begin{equation}
\nabla _{\mu }F^{\mu \nu }=0  \label{EMeq}
\end{equation}
where $G_{\mu \nu }$ is the Einstein tensor and $T_{\mu \nu }=2F_{%
\phantom{\lambda}{\mu}}^{\rho }F_{\rho \nu }-\frac{1}{2}F_{\rho \sigma
}F^{\rho \sigma }g_{\mu \nu }$ is the energy-momentum tensor of
electromagnetic field.

Since the second Lovelock term in Eq. (\ref{Act1}) is an Euler density in
four dimensions and has no contribution to the field equations in four or
less dimensional spacetimes, and we seek Taub-NUT/bolt solutions in even
dimensions, we therefore consider $(2k+2)$-dimensional spacetimes with $%
k\geq 2$. In constructing these metrics the idea is to regard the Taub-NUT
spacetime as a $U(1)$ fibration over a $2k$-dimensional base space endowed
with an Einstein-K\"{a}hler metric $g_{{\cal B}}$. Then the Euclidean
section of the $(2k+2)$-dimensional Taub-NUT spacetime can be written as:
\begin{equation}
ds^{2}=F(r)(d\tau +N{\cal A})^{2}+F^{-1}(r)dr^{2}+(r^{2}-N^{2})g_{{\cal B}}
\label{TN}
\end{equation}
where $\tau $ is the coordinate on the fibre $S^{1}$ and ${\cal A}$ has a
curvature $F=d{\cal A}$, which is proportional to some covariant constant
2-form. Here $N$ is the NUT charge and $F(r)$ is a function of $r$. The
solution will describe a `NUT' if the fixed point set of the $U(1)$ isometry
$\partial /\partial \tau $ (i.e. the points where $F(r)=0$) is less than $2k$%
-dimensional and a `bolt' if the fixed point set is $2k$-dimensional. We
assume the following form for the vector potential $A$ \cite{Awad2}
\begin{mathletters}
\begin{equation}
A=h(r)(d\tau +N{\cal A})  \label{pot}
\end{equation}
where ${\cal A}$ is the K\"{a}hler form of the base space ${\cal B}$ and $%
h(r)$ is an arbitrary function of $r$ which depends on the
dimension of the spacetime and is independent of the base space
${\cal B}$. In this paper we construct the Taub-NUT/bolt solutions
of Gauss-Bonnet gravity in the presence of electromagnetic field
and extend the two conjectures of Ref. \cite{Deh1} to the case of
electrically charged NUT solutions. These two conjectures were: 1)
For all non-extremal NUT solutions of Einstein gravity having no
curvature singularity at $r=N$, there exist NUT solutions in
Gauss-Bonnet gravity that contain these solutions in the limit
that the Gauss-Bonnet parameter $\alpha $\ vanishes. 2)
Gauss-Bonnet gravity has extremal NUT solutions whenever the base
space is a product of $2$-torii with at most one $2$-dimensional
space of positive curvature.

Here, we consider only the cases where all the factor spaces of ${\cal B}$
have zero or positive curvature. Thus, the base space ${\cal B}$ may be the
product of $2$-sphere $S^{2}$, $2$-torus $T^{2}$ or ${\Bbb CP}^{k}$. For
completeness, we give the $1$-forms and the metrics of these factor spaces.
The $1$-forms and metrics of $S^{2}$, $T^{2}$ and ${\Bbb CP}^{k}$ are
\end{mathletters}
\begin{eqnarray*}
{\cal A}_{i} &=&2\cos \theta _{i}d\phi _{i}, \\
d\Omega ^{2} &=&d\theta _{i}^{2}+\sin ^{2}\theta _{i}d\phi _{i}^{2}
\end{eqnarray*}
\begin{eqnarray*}
{\cal A}_{i} &=&2\eta _{i}d\zeta _{i} \\
d\Gamma _{i} &=&d\eta _{i}^{2}+d\zeta _{i}^{2}
\end{eqnarray*}
\begin{eqnarray}
{\cal A}_{k} &=&2(k+1)\sin ^{2}\xi _{k}\left( d\psi _{k}+\frac{1}{2k}{\cal A}%
_{k-1}\right)  \label{Ak} \\
d\Sigma _{k}^{2} &=&2(k+1)\left\{ d\xi _{k}^{2}+\sin ^{2}\xi _{k}\cos
^{2}\xi _{k}(d\psi _{k}+\frac{1}{2k}{\cal A}_{k-1})^{2}+\frac{1}{2k}\sin
^{2}\xi _{k}d\Sigma _{k-1}^{2}\right\}  \label{CPk}
\end{eqnarray}
respectively, where ${\cal A}_{k-1}$\ is the K\"{a}hler potential of ${\Bbb %
CP}^{k-1}$ \cite{Pop}. In Eqs. (\ref{Ak}) and (\ref{CPk}) $\xi _{k}$\ and $%
\psi _{k}$\ are the extra coordinates corresponding to ${\Bbb CP}^{k}$ with
respect to ${\Bbb CP}^{k-1}$. The metric ${\Bbb CP}^{k}$ is normalized such
that, Ricci tensor is equal to the metric, $R_{\mu \nu }=g_{\mu \nu }$. The $%
1$-form and the metric of ${\Bbb CP}^{1}$ are
\begin{eqnarray}
{\cal A}_{1} &=&4\sin ^{2}\xi _{1}d\psi _{1}  \label{A1} \\
d{\Sigma _{1}}^{2} &=&4\left( {d\xi _{1}}^{2}+\sin ^{2}\xi _{1}\cos ^{2}\xi
_{1}{d\psi _{1}}^{2}\right)  \label{CP1}
\end{eqnarray}

\section{Six-dimensional Solutions\label{6d}}

In this section we construct the six-dimensional Taub-NUT/bolt solutions of
the Gauss-Bonnet-Maxwell gravity. The base space ${\cal B}$ can be a $4$%
-dimensional space or a product of two $2$-dimensional spaces. The
electromagnetic field equation (\ref{EMeq}) for the metric (\ref{TN}) in six
dimensions is
\begin{equation}
(r^{2}-N^{2})^{2}h^{\prime \prime }(r)+4r(r^{2}-N^{2})h^{\prime
}(r)-8N^{2}h(r)=0  \label{max6eq}
\end{equation}
where through this paper the prime and double primes denote the first and
second derivative with respect to $r$ respectively. The solution of Eq. (\ref
{max6eq}) may be written as

\begin{equation}
h(r)=\frac{1}{(r^{2}-N^{2})^{2}}\{qr+V(r^{4}-6r^{2}N^{2}-3N^{4})\}
\label{solmaxeq6}
\end{equation}
where $q$ and $V$\ are two arbitrary constants which correspond to charge
and electric potential at infinity respectively.

To find the function $F(r)$, one may use any components of Eq. (\ref{Geq}).
The simplest equation is the $tt$ component of these equations which is
written in Sec. \ref{dd} for various base space in $2k+2$ dimensions. Here $%
k=2$, and we find that the function $F(r)$ for all the possible choices of
the base space ${\cal B}$ can be written in the form
\begin{eqnarray}
F(r) &=&\frac{(r^{2}-N^{2})^{2}}{12\alpha (r^{2}+N^{2})}\left( 1+\frac{%
p\alpha }{(r^{2}-N^{2})}-\sqrt{B(r)+C(r)}\right)   \nonumber \\
B(r) &=&1+\frac{4p\alpha N^{2}(r^{4}+6r^{2}N^{2}+N^{4})+12\alpha
mr(r^{2}+N^{2})}{(r^{2}-N^{2})^{4}}  \nonumber \\
&&+\frac{12\alpha \Lambda (r^{2}+N^{2})}{5(r^{2}-N^{2})^{4}}%
(r^{6}-5N^{2}r^{4}+15N^{4}r^{2}+5N^{6})  \nonumber \\
&&+\frac{3\alpha (r^{2}+N^{2})}{N^{3}(r^{2}-N^{2})^{6}}%
\{4N^{3}(3r^{2}-N^{2})q^{2}-128N^{5}r^{3}q V  \nonumber \\
&&+32N^{5}(r^{6}+15N^{2}r^{4}-9N^{4}r^{2}+9N^{6})V^{2}\}  \label{F6}
\end{eqnarray}
where $p$ is the sum of the dimensions of the curved factor spaces of ${\cal %
B}$, and the function $C(r)$\ depends on the choice of the base space ${\cal %
B}$. The function $C(r)$ for different base spaces are given in
the following table

\begin{center}
\begin{tabular}{ccc}
\hline\hline $\mathcal{B}$ &\hspace{.2cm} $p$ \hspace{.2cm}&
$(r^{2}-N^{2})^{4}C(r)/\alpha ^{2}$
\\ \hline
${{\mathbb{CP}^{2}}}$ &\hspace{.2cm} 4 \hspace{.2cm}& $-16(r^{4}+6r^{2}N^{2}+N^{4})+3D(r)$ \\
$S^{2}\times S^{2}$ &\hspace{.2cm} 4 \hspace{.2cm}& $-32(r^{4}+4r^{2}N^{2}+N^{4})-9D(r)$ \\
$T^{2}\times S^{2}$ &\hspace{.2cm} 2 \hspace{.2cm}& $4(r^{2}-N^{2})^{2}+9D(r)$ \\
$T^{2}\times T^{2}$ &\hspace{.2cm} 0 \hspace{.2cm}& $-9D(r)$ \\
\hline\hline
\end{tabular}\\
\end{center}

where

\begin{equation}
D(r)=\frac{r(r^{2}+N^{2})}{\alpha N^{3}}q^{2}
\end{equation}

One may note that the above solutions given in this section reduce to those
given in \cite{Deh1} as $q$ and $V$ vanish and reduce to the solutions
introduced in \cite{Awad2} as $\alpha $ goes to zero. Note that the
asymptotic behavior of these solutions for positive $\alpha $\ is locally
flat when $\Lambda $\ vanishes, locally dS for $\Lambda >0$ and locally AdS
for $\Lambda <0$\ provided $\left| \Lambda \right| <5/(12\alpha )$.

\subsection{Taub-NUT Solutions}

The solutions given in Eq. (\ref{F6}) describe NUT solutions, if (i) $%
F(r=N)=0$, (ii) $F^{\prime }(r=N)=1/(3N)$ and (iii) $h(r=N)=0$.
The first condition comes from the fact that all the extra
dimensions should collapse to zero at the fixed point set of
$\partial /\partial \tau $, the second one ensures that there is
no conical singularity with a smoothly closed fiber at $r=N$ and
the third one comes from the regularity of vector potential at
$r=N$. The last condition becomes

\begin{equation}
V\equiv V_{n}=\frac{q}{8N^{3}},  \label{Vq6}
\end{equation}
which is independent of the choice of the base space.

\begin{figure}[tbp]
\epsfxsize=10cm \centerline{\epsffile{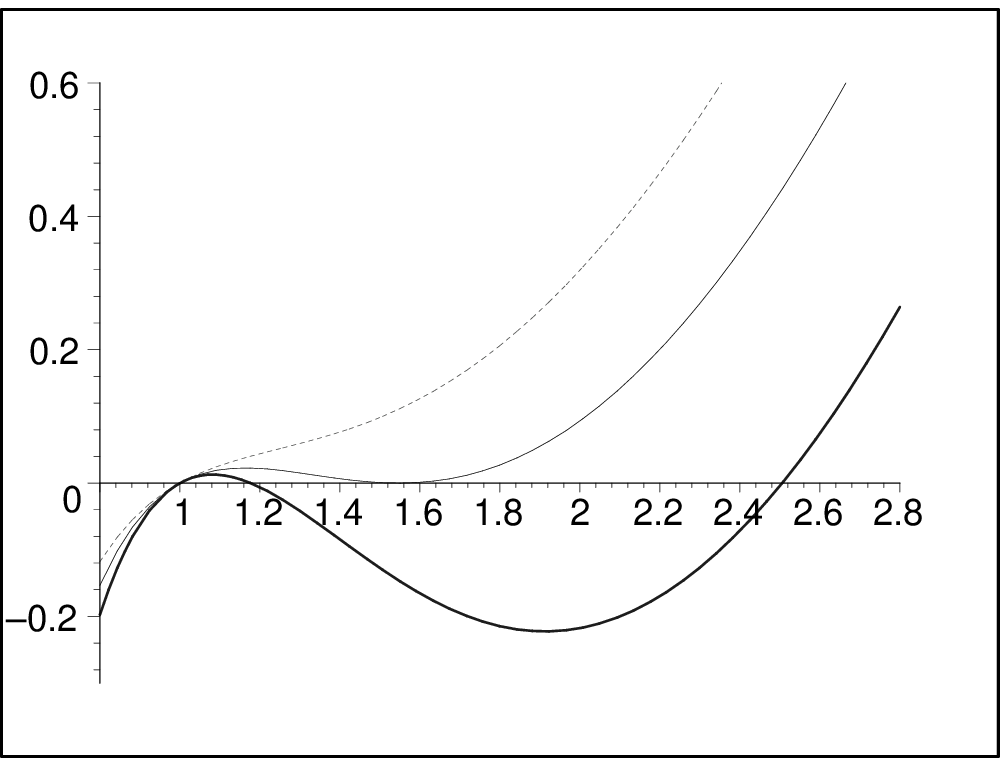}}
\caption{$F_{\mathrm{nut}}(r)$ versus $r$  with
$\mathcal{B}=\mathbb{CP}^{2}$ for $N=1$, $\alpha=0.1$,
$\Lambda=-2$, and $q=q_{\mathrm{crit}}=24.815$ (continuous line),
$q<q_{\mathrm{crit}}$ (dotted line) and $q>q_{\mathrm{crit}}$
(bold line).} \label{Fig6cp}
\end{figure}

Using the first two conditions with Eq. (\ref{Vq6}), one finds that
Gauss-Bonnet-Maxwell gravity in six dimensions admits NUT solutions with a $%
{\Bbb CP}^{2}$ base space when the mass parameter is fixed to be
\begin{equation}
m_{n}=-\frac{16}{15}N(3\Lambda N^{4}+5N^{2}-5\alpha )-\frac{3}{4}\frac{q^{2}%
}{N^{3}}  \label{mn6cp}
\end{equation}
provided the charged parameter $q$ is less than a critical value $q_{{\rm %
crit}}$. This condition on $q$ comes up from the fact that the horizon at $%
r=N$ may not be the event horizon. Indeed for $q\geq q_{{\rm crit}}$ the
event horizon located at $r>N$. To find $q_{{\rm crit}}$ we proceeds as
follows. We define the function $g_{{\rm nut}}(r)$ as the numerator of $F_{%
{\rm nut}}(r)=F(V=V_{n},m=m_{n},r)/(r-N)$ which is positive at $r=N$, and
solve the system of two equations
\begin{equation}
\left\{
\begin{array}{l}
g_{{\rm nut}}(r) =0 \\
g_{{\rm nut}}^{\prime }(r) =0 \label{gnut}
\end{array}
\right.
\end{equation}
for the unknown $q$ and $r$. The $q$ obtained by this method is the critical
value $q_{{\rm crit}}$. To be more clear, we first obtain $q_{{\rm crit}}$
for the case of $\Lambda =\alpha =0$. The system of two equations (\ref{gnut}%
) becomes
\[
\left\{
\begin{array}{l}
16N^{4}(r+3N)(r+N)^{2}-3(r-N)q^{2} =0, \\
16N^{4}(3r+7N)(r+N)-3q^{2} =0
\end{array}
\right.
\]
with the following solution for $q$
\begin{equation}
q_{{\rm crit}}=\left\{ \frac{32}{3}(11+5\sqrt{5})\right\} ^{1/2}N^{3}
\end{equation}
For arbitrary values of $\Lambda $ and $\alpha $, one may find the critical
value of $q$ numerically. For $\alpha =0.1${\bf , }$\Lambda =-2$ and $N=1$
the critical value of charge which is obtained by solving the system of two
equations (\ref{gnut}) is $q_{{\rm crit}}=24.815$. This can be seen in Fig.
\ref{Fig6cp} which shows the function $F_{{\rm nut}}(r)$ as a function of $r$
for various values of $q$ including $q=q_{{\rm crit}}$.

As in the case of solutions of Gauss-Bonnet gravity in the absence of
electromagnetic field, the solution with base space ${\cal B}=S^{2}\times
S^{2}$ does not satisfy the conditions of NUT solutions. Computation of the
Kretschmann scalar at $r=N$ for the solutions in six dimensions shows that
the spacetime with ${\cal B}=S^{2}\times S^{2}$ has a curvature singularity
at $r=N$ in Einstein gravity, while the spacetime with ${\cal B}={\Bbb CP}%
^{2}$ has no curvature singularity at $r=N$. Thus, the conjecture given in
\cite{Deh1} is confirmed even in the presence of electromagnetic field.
Indeed, we have non-extreme NUT solutions in $6$ dimensions with non-trivial
fibration when the $4$-dimensional base space is chosen to be ${\Bbb CP}^{2}$%
.

\begin{figure}[tbp]
\epsfxsize=10cm \centerline{\epsffile{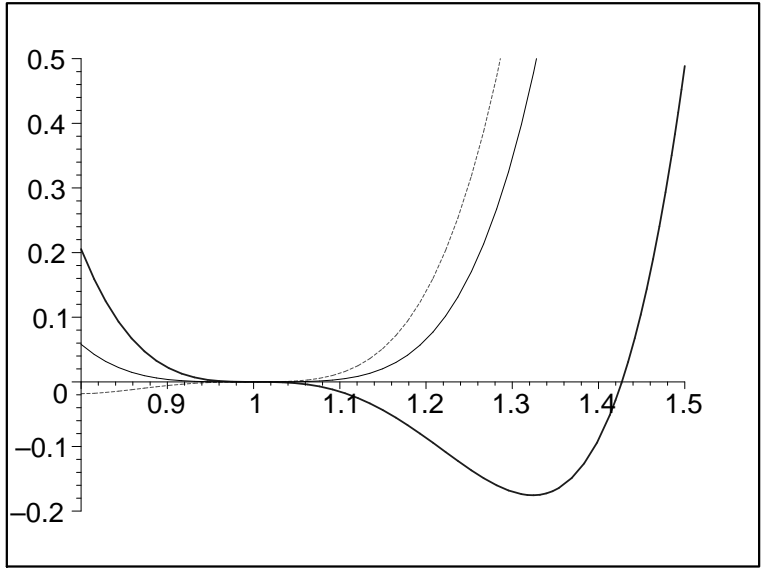}}
\caption{$F_{\mathrm{nut}}(r)$ versus $r$ with
$\mathcal{B}_{B}=T^{2}\times T^{2}$ for $N=1$, $\alpha=0.1$,
$\Lambda=-2$, and $q_{\mathrm{crit}}=11.322$ (continuous line),
$q<q_{\mathrm{crit}}$ (dotted line) and $q>q_{\mathrm{crit}}$
(bold line).} \label{Fig6TT}
\end{figure}

\begin{figure}[h]
\epsfxsize=10cm \centerline{\epsffile{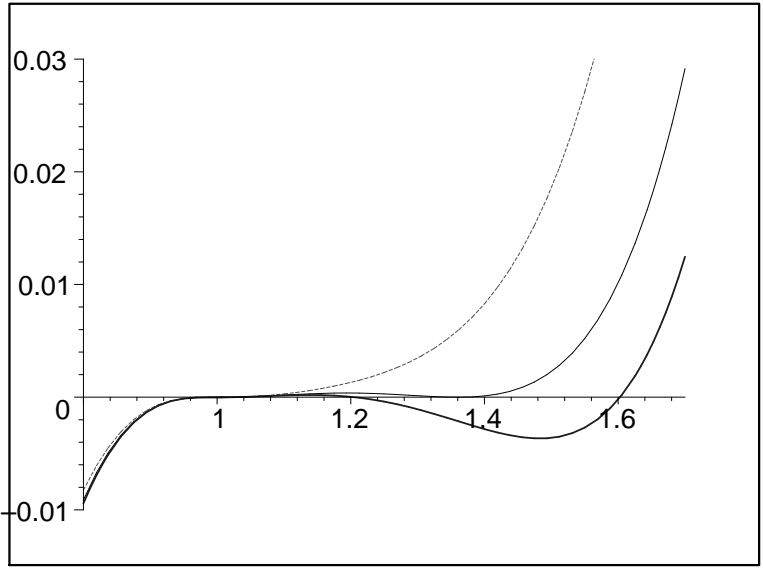}}
\caption{$F_{\mathrm{nut}}(r)$ versus $r$ with
$\mathcal{B}_{B}=T^{2}\times S^{2}$ for $N=1$, $\alpha=0.1$,
$\Lambda=-2$, and $q_{\mathrm{crit}}=19.973$ (continuous line),
$q<q_{\mathrm{crit}}$ (dotted line) and $q>q_{\mathrm{crit}}$
(bold line).} \label{Fig6TS}
\end{figure}

On the other hand, the solutions with ${\cal B}=T^{2}\times T^{2}={\cal B}%
_{A}$ and ${\cal B}=T^{2}\times S^{2}={\cal B}_{B}$ are extermal NUT
solution provided the charge parameter is less than the critical value $q_{%
{\rm crit}}$ and the mass parameter is fixed to be
\begin{eqnarray}
m_{n}^{A} &=&-\frac{16}{5}\Lambda N^{5}-\frac{3}{4}\frac{q^{2}}{N^{3}}, \\
m_{n}^{B} &=&-\frac{8}{15}N^{3}(6\Lambda N^{2}+5)-\frac{3}{4}\frac{q^{2}}{%
N^{3}}
\end{eqnarray}
Indeed for these two cases $F^{\prime }(r=N)=0$, and therefore the NUT
solutions should be regarded as extremal solutions. As in the case of
non-extreme NUT solution, the critical value $q_{{\rm crit}}$ depends on $%
\alpha $, $\Lambda $ and $N$, and it is not easy to give an analytic
expression for it. The critical value of $q$ for ${\cal B}_{A}=T^{2}\times
T^{2}$ and ${\cal B}_{B}=T^{2}\times S^{2}$ may be found by solving the
system of two equations (\ref{gnut}). See Figs. \ref{Fig6TT} and \ref{Fig6TS}
for more details.


Computing the Kretschmann scalar, we find that there is a curvature
singularity at $r=N$ \ for the spacetime with ${\cal B}={\cal B}_{B}$, while
the spacetime with ${\cal B}_{A}$ has no curvature singularity at $r=N$.
Thus, the second conjecture of Ref. \cite{Deh1} can be extended to the case
of Gauss-Bonnet gravity in the presence of electromagnetic field. Indeed,
when the base space has at most one two dimensional curved space as one of
its factor spaces, then Gauss-Bonnet-Maxwell gravity admits an extreme NUT
solution even though there exists a curvature singularity at $r=N $. As in
the case of uncharged solutions of Gauss-Bonnet gravity, the extreme NUT
solution for the base space $T^{2}\times T^{2}$ in the absence of
cosmological constant ($\Lambda =0$) has no horizon and the singularity is
naked.

\subsection{Taub-Bolt Solutions}

The conditions for having a regular bolt solution are (i) $F(r=r_{b})=0$,
(ii)$\ F^{\prime }(r_{b})=1/(3N)$ and (iii)

\begin{equation}
V\equiv V_{b}=-\frac{qr_{b}}{r_{b}^{4}-6N^{2}r_{b}^{2}-3N^{4}}
\label{Vbolt6}
\end{equation}
with $r_{b}>N$. Condition (ii), which again follows from the fact that we
want to avoid a conical singularity at the bolt, together with the fact that
the period of $\tau $ will still be $12\pi N$ and $V=V_{b}$, gives the
following equation for $r_{b}$
\[
3N\Lambda {r_{b}}^{4}+2{r_{b}}^{3}-6N(\Lambda N^{2}+1){r_{b}}%
^{2}-2(N^{2}-4\alpha )r_{b}+3\Lambda N^{5}+6N^{3}-\zeta \alpha N-27N(\frac{%
V_{b}}{r_{b}})^{2}(r_{b}^{2}-N^{2})^{2}=0
\]
where $\zeta $ is $8$ and $12$ for the base spaces ${\Bbb CP}^{2}$ and $%
S^{2}\times S^{2}$ respectively.

Next we consider the Taub-bolt solutions for ${\cal B}=T^{2}\times S^{2}$
and ${\cal B}=T^{2}\times T^{2}$. Euclidean regularity at the bolt requires
the period of $\tau $ to be
\begin{equation}
\beta =\frac{8\pi r_{b}^{3}(r_{b}^{2}-N^{2}+2\alpha )}{%
r_{b}^{2}(r_{b}^{2}-N^{2})[1-\Lambda
(r_{b}^{2}-N^{2})]+9V_{b}^{2}(r_{b}^{2}-N^{2})^{2}}  \label{betst6}
\end{equation}
for ${\cal B}=T^{2}\times S^{2}$, and

\begin{equation}
\beta =-\frac{8\pi r_{b}^{3}}{(r_{b}^{2}-N^{2})(\Lambda r_{b}^{2}-9V_{b}^{2})%
}  \label{bettt6}
\end{equation}
for ${\cal B}=T^{2}\times T^{2}$. As $r_{b}$ varies from $N$ to
infinity, one covers the whole temperature range from $0$ to
$\infty $, and therefore we have non-extreme bolt solutions.
Indeed, the fibration in the latter case is trivial: there are no
Misner strings. The boundary has trivial topology and therefore
the Euclidean time period $\beta$ will not be fixed, as it was in
the ${\cal B}={\Bbb CP}^{2}$ case, by the value of the NUT
parameter \cite{Myers2}. Again, as in the case of uncharged
solution \cite{Deh1}, there is no bolt solution with ${\cal
B}=T^{2}\times T^{2}$ in the absence of cosmological constant.

\section{Eight-dimensional Solutions\label{8d}}

In eight dimensions there are more possibilities for the base space ${\cal B}
$. It can be a $6$-dimensional space, a product of three $2$-dimensional
spaces, or the product of a $4$-dimensional space with a $2$-dimensional
one. We first consider the differential equation for vector potential (\ref
{pot}). Equation (\ref{EMeq}) has the same form for any base space ${\cal B}$
as

\begin{equation}
(r^{2}-N^{2})^{2}h^{\prime \prime }(r)+6r(r^{2}-N^{2})h^{\prime
}(r)-12N^{2}h(r)=0  \label{max8eq}
\end{equation}
with the solution

\begin{equation}
h(r)=\frac{1}{(r^{2}-N^{2})^{3}}\left[
qr+V(r^{6}-5N^{2}r^{4}+15N^{4}r^{2}+5N^{6})\right]  \label{solmaxeq8}
\end{equation}
where $V$ and $q$\ are two arbitrary constants which correspond to electric
potential at infinity \ and electric charge respectively.

For any base space, the form of the function $F(r)$ is

\begin{eqnarray}
F(r) &=&\frac{(r^{2}-N^{2})^{2}}{8\alpha (5r^{2}+3N^{2})}\left( 1+\frac{%
4p\alpha }{3(r^{2}-N^{2})}-\sqrt{B(r)+C(r)}\right)  \nonumber \\
B(r) &=&1-\frac{16\alpha mr\left( 5r^{2}+3N^{2}\right) }{3(r^{2}-N^{2})^{5}}+%
\frac{16p\alpha N^{2}}{15(r^{2}-N^{2})^{5}}%
(r^{6}-15N^{2}r^{4}-45N^{4}r^{2}-5N^{6})  \nonumber \\
&&+\frac{16\alpha \Lambda \left( 5r^{2}+3N^{2}\right) }{105(r^{2}-N^{2})^{5}}%
(5r^{8}-28N^{2}r^{6}+70N^{4}r^{4}-140N^{6}r^{2}-35N^{8})  \nonumber \\
&&+\frac{4\alpha \left( 5r^{2}+3N^{2}\right) }{9N^{5}(r^{2}-N^{2})^{8}}%
\{12N^{5}(5r^{2}-N^{2})q^{2}-384N^{7}r^{3}(r^{2}-5N^{2})qV  \nonumber \\
&&+48N^{7}(r^{10}-25N^{2}r^{8}-70N^{4}r^{6}+350N^{6}r^{4}-75N^{8}r^{2}+75N^{10})V^{2}\}
\label{F8}
\end{eqnarray}
where $p$\ is again the dimension of the curved factor spaces of ${\cal B} $%
, and the function $C(r)$\ depends on the choice of the base space. The
function $C(r)$ for various base spaces are\newline

\begin{center}
\begin{tabular}{ccc}
\hline\hline ${\cal B}$ & \hspace{.3cm} $p$ \hspace{.2cm} & $(r^{2}-N^{2})^{5}C(r)/\alpha ^{2}$ \\
\hline ${\,{{\Bbb CP}^{3}}}$ & \hspace{.3cm} 6 \hspace{.2cm} &
$-16(r^{6}-15N^{2}r^{4}-45N^{4}r^{2}-5N^{6})-20D(r) $\\
$S^{2}\times {\Bbb CP}^{2}$ & \hspace{.2cm} 6 \hspace{.2cm}& $-\frac{64}{27}%
(13r^{6}-135N^{2}r^{4}-345N^{4}r^{2}-45N^{6})+60D(r)$ \\
$S^{2}\times S^{2}\times S^{2}$ & \hspace{.2cm} 6 \hspace{.2cm}& $-\frac{128}{3}%
(r^{6}-9N^{2}r^{4}-21N^{4}r^{2}-3N^{6})+20D(r)$ \\
$T^{2}\times {\Bbb CP}^{2}$ & \hspace{.2cm} 4 \hspace{.2cm}&$\frac{128}{27}%
(r^{6}+9N^{2}r^{4}+51N^{4}r^{2}+3N^{6})-60D(r)$ \\
$T^{2}\times S^{2}\times S^{2}$ & \hspace{.2cm} 4 \hspace{.2cm}& $-\frac{64}{9}%
(r^{6}-15N^{2}r^{4}-45N^{4}r^{2}-5N^{6})-60D(r)$ \\
$T^{2}\times T^{2}\times S^{2}$ & \hspace{.2cm} 2 \hspace{.2cm}& $\frac{64}{9}(r^{2}-N^{2})^{3}-20D(r)$ \\
$T^{2}\times T^{2}\times T^{2}$ & \hspace{.2cm} 0
\hspace{.2cm}&--$20D(r)$ \\ \hline\hline
\end{tabular}
\\[0pt]
\end{center}

where

\begin{equation}
D(r)=\frac{r(5r^{2}+3N^{2})}{9\alpha N^{5}}q^{2}  \label{D8}
\end{equation}

One may note that the asymptotic behavior of all of these solutions is
locally AdS for $\Lambda <0$ provided $\left| \Lambda \right| <21/(80\alpha
) $, locally dS for $\Lambda >0$ and locally flat for $\Lambda =0$. Note
that all the different $F(r)$'s given in this section have the same form as $%
\alpha $ goes to zero. Also, one may note that these solutions reduce to the
solutions of Gauss-Bonnet gravity \cite{Deh1} when $q$ and $V$ vanish.

\subsection{Taub-NUT Solutions}

As in the case of six-dimensional spacetimes, the solutions given in Eq. (%
\ref{F8}) describe NUT solutions, if (i) $F(r=N)=0$, (ii) $F^{\prime
}(r=N)=1/(4N)$, (iii) $h(r=N)=0$ and (iv) $q<q_{{\rm crit}}$, where $q_{{\rm %
crit}}$ is the solution of the system of two equations (\ref{gnut}). Using
the third conditions which comes from the regularity of vector potential at $%
r=N$ \ gives

\begin{equation}
V\equiv V_{n}=-\frac{q}{16N^{5}},  \label{Vn8}
\end{equation}
It is easy to show that Gauss-Bonnet-Maxwell gravity in eight dimensions
admits non-extreme NUT solutions only when the base space is chosen to be $%
{\Bbb CP}^{3}$. The conditions for a nonsingular NUT solution are satisfied
provided the mass parameter is fixed to be
\begin{equation}
m_{n}=-\frac{8N^{3}}{105}(16\Lambda N^{4}+42N^{2}-105\alpha )-\frac{5}{12}%
\frac{q^{2}}{N^{5}}  \label{mcp3}
\end{equation}

On the other hand, the solutions with ${\cal B}=T^{2}\times T^{2}\times
T^{2}={\cal B}_{A}$ and ${\cal B}=T^{2}\times T^{2}\times S^{2}={\cal B}_{B}$
are extermal NUT solutions provided the mass parameter is

\begin{eqnarray}
m_{n}^{A} &=&-\frac{128}{105}\Lambda N^{7}-\frac{5}{12}\frac{q^{2}}{N^{5}},
\label{mttt8} \\
m_{n}^{B} &=&-\frac{16N^{5}}{105}(8\Lambda N^{2}+7)-\frac{5}{12}\frac{q^{2}}{%
N^{5}}  \label{mstt8}
\end{eqnarray}
These results for eight-dimensional Gauss-Bonnet gravity are consistent with
the conjectures of Ref. \cite{Deh1}. Again, one may note that the former
extremal NUT solution does not have a curvature singularity at $r=N$ whereas
the latter does.

\subsection{Taub-Bolt Solutions}

The conditions for having a regular bolt solution are $F(r=r_{b})=0$, $%
F^{\prime }(r_{b})=1/(4N)$ and

\begin{equation}
V\equiv V_{b}=-\frac{qr_{b}}{%
r_{b}^{6}-5N^{2}r_{b}^{4}+15N^{4}r_{b}^{2}+5N^{6}}  \label{Vb8}
\end{equation}
with $r_{b}>N$. The second condition again follows from the fact that we
want to avoid a conical singularity at the bolt, together with the fact that
the period of $\tau $ will still be $16\pi N$. Now applying these conditions
for the curved base spaces gives the following equation for $r_{b}$

\[
4N\Lambda {r_{b}}^{4}+3{r_{b}}^{3}-4N(3+2\Lambda N^{2}){r_{b}}^{2}+3(8\alpha
-N^{2})r_{b}+4N(\Lambda N^{4}+3N^{2}-\zeta \alpha )-100N(\frac{V_{b}}{r_{b}}%
)^{2}(r_{b}^{2}-N^{2})^{2}=0
\]
where $\zeta $ is $9$, $32/3$ and $12$ for the base spaces ${\Bbb CP}^{3}$, $%
S^{2}\times {\Bbb CP}^{2}$ and $S^{2}\times S^{2}\times S^{2}$ respectively.

For the case of ${\cal B}=T^{2}\times T^{2}\times T^{2}$ and ${\cal B}%
=T^{2}\times T^{2}\times S^{2}$, Euclidean regularity at the bolt requires
the period of $\tau $ to be
\begin{equation}
\beta =-\frac{12\pi r_{b}^{3}}{({r_{b}}^{2}-N^{2})(\Lambda {r_{b}}%
^{2}-25V_{b}^{2})}
\end{equation}

and
\begin{equation}
\beta =\frac{4\pi r_{b}^{3}(3{r_{b}}^{2}-3N^{2}+8\alpha )}{{r_{b}}^{2}({r_{b}%
}^{2}-N^{2})[1-\Lambda ({r_{b}}^{2}-N^{2})]+25V_{b}^{2}({r_{b}}%
^{2}-N^{2})^{2}}
\end{equation}
respectively. As $r_{b}$ varies from $N$ to infinity, one covers the whole
temperature range from $0$ to $\infty $, and therefore one can have bolt
solutions. Again, one may note that there is no asymptotic locally flat
black hole solutions with base space ${\cal B}=T^{2}\times T^{2}\times T^{2}$%
.

\section{Ten-dimensional Solutions\label{10d}}

In ten dimensions there are more possibilities for the base space ${\cal B}$%
. It can be an $8$-dimensional space, the product of a $6$-dimensional space
with a $2$-dimensional one, a product of two $4$-dimensional spaces, a
product of a $4$-dimensional space with two $2$-dimensional spaces, or the
product of four $2$-dimensional spaces. Substituting the vector potential (%
\ref{pot}) in source free Maxwell equation (\ref{EMeq}), for a
ten-dimensional spacetime of the form given in Eq. (\ref{TN}) with an
arbitrary base space ${\cal B}$ , one obtains

\begin{equation}
(r^{2}-N^{2})^{2}h^{\prime \prime }(r)+8r(r^{2}-N^{2})h^{\prime
}(r)-16N^{2}h(r)=0  \label{max10eq}
\end{equation}
with the solution

\begin{equation}
h(r)=\frac{1}{5(r^{2}-N^{2})^{4}}%
\{5qr+V(5r^{8}-28N^{2}r^{6}+70N^{4}r^{4}-140N^{6}r^{2}-35N^{8})\}
\label{solmaxeq10}
\end{equation}
where $V$\ and $q$ are two arbitrary constants which correspond to electric
potential at infinity and electric charge respectively.

The form of the function $F(r)$ for any base space ${\cal B}$\ may be
written as

\begin{eqnarray}
F(r) &=&\frac{(r^{2}-N^{2})^{2}}{12\alpha (7r^{2}+3N^{2})}\left( 1+\frac{%
3p\alpha }{2(r^{2}-N^{2})}-\sqrt{B(r)+C(r)}\right) ,  \nonumber \\
B(r) &=&1+\frac{36\alpha mr\left( 7r^{2}+3N^{2}\right) }{(r^{2}-N^{2})^{6}}+%
\frac{6p\alpha N^{2}}{35(r^{2}-N^{2})^{6}}%
(3r^{8}-28N^{2}r^{6}+210N^{4}r^{4}+420N^{6}r^{2}+35N^{8})  \nonumber \\
&&+\frac{2\alpha \Lambda \left( 7r^{2}+3N^{2}\right) }{21(r^{2}-N^{2})^{6}}%
(7r^{10}-45N^{2}r^{8}-126N^{4}r^{6}-210N^{6}r^{4}+315N^{8}r^{2}+63N^{10})
\nonumber \\
&&+\frac{3\alpha \left( 7r^{2}+3N^{2}\right) }{32N^{7}(r^{2}-N^{2})^{10}}%
\{64N^{7}(7r^{2}-N^{2})q^{2}-4096N^{9}r^{3}(35N^{4}-14N^{2}r^{2}+3r^{4})qV+
\nonumber \\
&&1024N^{9}(5r^{14}-77N^{2}r^{12}+861N^{4}r^{10}-525N^{6}r^{8}-5145N^{8}r^{6}+11025N^{10}r^{4}
\nonumber \\
&&-1225N^{12}r^{2}+1225N^{14})V^{2}\}
\end{eqnarray}
where $p$ is the dimensionality of the curved portion of the base space, and
the function $C(r)$ depends on the choice of the base space ${\cal B}$. The
function $C(r)$ for different base spaces are listed in the following table%
\newline

\begin{center}
\begin{tabular}{ccc}
\hline\hline ${\cal B}$ & \hspace{.2cm} $p$ \hspace{.2cm} &
$(r^{2}-N^{2})^{6}C(r)/\alpha ^{2}$ \\ \hline
${{\Bbb CP}^{4}}$ & \hspace{.2cm} 8 \hspace{.2cm} & $-\frac{144}{25}%
(3r^{8}-28N^{2}r^{6}+210N^{4}r^{4}+420N^{6}r^{2}+35N^{8})-21D(r)$ \\
${\Bbb CP}^{2}\times {\Bbb CP}^{2}$ & \hspace{.2cm} 8 \hspace{.2cm} & $-\frac{16}{5}%
(11r^{8}-76N^{2}r^{6}+450N^{4}r^{4}+820N^{6}r^{2}+75N^{8})-630D(r)$ \\
$S^{2}\times {\Bbb CP}^{3}$ & \hspace{.2cm} 8 \hspace{.2cm} & $-\frac{18}{5}%
(9r^{8}-64N^{2}r^{6}+390N^{4}r^{4}+720N^{6}r^{2}+65N^{8})+630D(r)$ \\
$S^{2}\times S^{2}\times {\Bbb CP}^{2}$ & \hspace{.2cm} 8 \hspace{.2cm} & $-\frac{8}{5}%
(29r^{8}-184N^{2}r^{6}+990N^{4}r^{4}+1720N^{6}r^{2}+165N^{8})-630D(r)$ \\
$S^{2}\times S^{2}\times S^{2}\times S^{2}$ & \hspace{.2cm} 8 \hspace{.2cm} & -$\frac{285}{5}%
(r^{8}-6N^{2}r^{6}+300N^{4}r^{4}+50N^{6}r^{2}+5N^{8})-630D(r)$ \\
$T^{2}\times {\Bbb CP}^{3}$ & \hspace{.2cm} 6 \hspace{.2cm} & $\frac{27}{5}%
(r^{8}+4N^{2}r^{6}-90N^{4}r^{4}-220N^{6}r^{2}-15N^{8})-630D(r)$ \\
$T^{2}\times S^{2}\times {\Bbb CP}^{2}$ & \hspace{.2cm} 6 \hspace{.2cm} & $-\frac{1}{5}%
(43r^{8}-428N^{2}r^{6}+3330N^{4}r^{4}+6740N^{6}r^{2}+555N^{8})-630D(r)$ \\
$T^{2}\times S^{2}\times S^{2}\times S^{2}$ & \hspace{.2cm} 6 \hspace{.2cm} & $-\frac{9}{5}%
(11r^{8}-76N^{2}r^{6}+450N^{4}r^{4}+820N^{6}r^{2}+75N^{8})-630D(r)$ \\
$T^{2}\times T^{2}\times {\Bbb CP}^{2}$ & \hspace{.2cm} 4 \hspace{.2cm} & $\frac{4}{5}%
(17r^{8}-52N^{2}r^{6}-90N^{4}r^{4}-500N^{6}r^{2}-15N^{8})-630D(r)$ \\
$T^{2}\times T^{2}\times S^{2}\times S^{2}$ & \hspace{.2cm} 4 \hspace{.2cm} & $\frac{12}{5}%
(r^{8}+4N^{2}r^{6}-90N^{4}r^{4}-220N^{6}r^{2}-15N^{8})-630D(r)$ \\
$T^{2}\times T^{2}\times T^{2}\times S^{2}$ & \hspace{.2cm} 2
\hspace{.2cm} & $9(r^{2}-N^{2})^{4}+105D(r)$
\\
$T^{2}\times T^{2}\times T^{2}\times T^{2}$ & \hspace{.2cm} 0
\hspace{.2cm} & $105D(r)$ \\ \hline\hline
\end{tabular}
\\[0pt]
\end{center}

where

\begin{equation}
D(r)=\frac{r(7r^{2}+3N^{2})}{32\alpha N^{7}}q^{2}
\end{equation}

Note that the asymptotic behavior of all of these solutions is locally AdS
for $\Lambda <0$ provided $\left| \Lambda \right| <9/(42\alpha )$, locally
dS for $\Lambda >0$ and locally flat for $\Lambda =0$. As with the $6$ and $%
8 $ dimensional cases, all the different $F(r)$'s have the same form as $%
\alpha $ goes to zero. Also, one may note that these solutions reduce to
those given in \cite{Deh1} when $q$ and $V$ vanish.

\subsection{Taub-NUT Solutions}

In order to have NUT solutions, the four conditions (i) $F(r=N)=0$, (ii) $%
F^{\prime }(r=N)=1/(5N)$, (iii) the regularity of vector potential at $r=N$,

\begin{equation}
V\equiv V_{n}=\frac{5q}{128N^{7}},  \label{VN10}
\end{equation}
and (iv) the restriction on $q<q_{{\rm crit}}$, where $q_{{\rm crit}}$ is
the solution of the system of two equations (\ref{gnut}) should be
satisfied. Using these four condition, we find that Gauss-Bonnet gravity in
ten dimensions admits non-extreme NUT solutions only when the base space is
chosen to be ${\Bbb CP}^{4}$, provided the mass parameter is fixed to be

\begin{equation}
m_{n}=-\frac{128N^{5}}{4725}(25\Lambda N^{4}+90N^{2}-378\alpha )-\frac{35}{%
128}\frac{q^{2}}{N^{7}}
\end{equation}

On the other hand, the solutions with ${\cal B}=T^{2}\times T^{2}\times
T^{2}\times T^{2}={\cal B}_{A}$ and ${\cal B}=T^{2}\times T^{2}\times
T^{2}\times S^{2}={\cal B}_{B}$ are extermal NUT solution provided the mass
parameter is
\begin{eqnarray}
m_{n}^{A} &=&-\frac{128}{189}\Lambda N^{9}-\frac{35}{128}\frac{q^{2}}{N^{7}},
\label{mttt10} \\
m_{n}^{B} &=&-\frac{64N^{7}}{945}(10\Lambda N^{2}+9)-\frac{35}{128}\frac{%
q^{2}}{N^{7}}  \label{mstt10}
\end{eqnarray}
and $q<q_{{\rm crit}}$. It is also straightforward to show that the former
extremal NUT solution has no curvature singularity at $r=N$, whereas the
latter has. These results in ten-dimensions shows that the conjectures of
Ref. \cite{Deh1} may be extended to the case of Gauss-Bonnet-Maxwell gravity.

\subsection{Taub-Bolt Solutions}

Now applying the conditions for having a regular bolt solution $F(r=r_{b})=0$%
, $F^{\prime }(r_{b})=1/(5N)$ with $r_{b}>N$ and
\[
V\equiv V_{b}=-\frac{5qr_{b}}{%
5r_{b}^{8}-28N^{2}r_{b}^{6}+70N^{4}r_{b}^{4}-140N^{6}r_{b}^{2}-35N^{8}}
\]
for the curved base spaces gives the following equation for $r_{b}$%
\begin{eqnarray*}
&& 5N\Lambda {r_{b}}^{4}+4{r_{b}}^{3}-10N(2+\Lambda N^{2}){r_{b}}%
^{2}+4(12\alpha -N^{2})r_{b} \\
&&+N(5\Lambda N^{4}+20N^{2}-\zeta \alpha )-245N(\frac{V_{b}}{r_{b}}%
)^{2}(r_{b}^{2}-N^{2})^{2}=0
\end{eqnarray*}
where $\zeta$ is equal to 96, 105, 320/3, 340/3 and 120 for the base spaces $%
{\Bbb CP}^{4}$, $S^{2}\times {\Bbb CP}^{3}$, ${\Bbb CP}^{2}\times {\Bbb CP}%
^{2}$, $S^{2}\times S^{2}\times {\Bbb CP}^{2}$ and $S^{2}\times S^{2}\times
S^{2}\times S^{2}$ respectively.

For the case of ${\cal B}=T^{2}\times T^{2}\times T^{2}\times T^{2}$ and $%
{\cal B}=S^{2}\times T^{2}\times T^{2}\times T^{2}$, Euclidean regularity at
the bolt requires the period of $\tau $ to be
\begin{equation}
\beta =-\frac{16\pi r_{b}^{3}}{({r_{b}}^{2}-N^{2})(\Lambda {r_{b}}%
^{2}-49V_{b}^{2})}
\end{equation}

and
\begin{equation}
\beta =\frac{16\pi r_{b}^{3}({r_{b}}^{2}-N^{2}+3\alpha )}{{r_{b}}^{2}({r_{b}}%
^{2}-N^{2})[1-\Lambda ({r_{b}}^{2}-N^{2})]+49V_{b}^{2}({r_{b}}^{2}-N^{2})^{2}%
}
\end{equation}
respectively. As $r_{b}$ varies from $N$ to infinity, one covers the whole
temperature range from $0$ to $\infty $, and therefore one can have bolt
solutions. Again, one may note that for the case of an asymptotic locally
flat solution with base space ${\cal B}=T^{2}\times T^{2}\times T^{2}\times
T^{2}$, there is no black hole solution.

\section{The $(2k+2)$-dimensional Taub-NUT\label{dd} solutions}

In this section we present the $(2k+2)$-dimensional solution of
Gauss-Bonnet-Maxwell gravity. The electromagnetic field equation (\ref{EMeq}%
) for the metric (\ref{TN}) in $2k+2$ dimensions is
\begin{equation}
(r^{2}-N^{2})^{2}h^{\prime \prime }(r)+2kr(r^{2}-N^{2})h^{\prime
}(r)-4kN^{2}h(r)=0  \label{EMd}
\end{equation}

The solution of Eq. (\ref{EMd}) may be expressed in terms of hypergeometric
function $_{2}F_{1}([a,b],[c],z)$\ in a compact form. The result is
\begin{equation}
h(r)=\frac{1}{(r^{2}-N^{2})^{k}}\left(
qr-(-1)^{k}(2k-1)VN^{2k}\,_{2}F_{1}\left( \left[ -\frac{1}{2},-k\right] ,%
\left[ \frac{1}{2}\right] ,\frac{r^{2}}{N^{2}}\right) \right)  \label{Hrd}
\end{equation}
where $V$\ and $q$ are two arbitrary constants which correspond to electric
potential at infinity and charge respectively.

Here, we consider only those cases which Gauss-Bonnet-Maxwell gravity admits
NUT solutions, leaving out the other cases which one has only bolt solution
for reasons of economy. There are three cases which we have NUT solutions in
$2k+2$ dimensions.

The only case which Gauss-Bonnet gravity admits non-extreme NUT solution in $%
2k+2$ dimensions is when the base space is ${\cal B}={\Bbb CP}^{k}$. To find
the function $F(r)$, one may use any components of Eq. (\ref{Geq}). The
simplest equation is the $tt$ component of these equations which can be
written as
\begin{equation}
\Omega _{1}rF^{\prime }(r)+\Omega _{2}F^{2}(r)+\Omega _{3}F(r)+\Omega _{4}=
\left[ h^{\prime \prime }(r)\right] ^{2}+\frac{4kN^{2}}{(r^{2}-N^{2})^{2}}%
h^{2}(r),  \label{Eqd}
\end{equation}
where $h(r)$ is given in Eq. (\ref{Hrd}) and $\Omega _{1}$ to $\Omega _{4}$
are
\begin{eqnarray}
\Omega _{1} &=&-\alpha \left( (2k-1)r^{2}+3N^{2}\right)
F(r)+(r^{2}-N^{2})\left( \alpha +\frac{r^{2}-N^{2}}{4(k-1)}\right) ,
\nonumber \\
\Omega _{2} &=&-\frac{\alpha }{2(r^{2}-N^{2})}\left\{
(2k-1)(2k-3)r^{4}+2(2k-7)N^{2}r^{2}+3N^{4}\right\} ,  \nonumber \\
\Omega _{3} &=&-\frac{(r^{2}-N^{2})[(2k-1)r^{2}+N^{2}]}{4(k-1)}+\alpha
\left( (2k-3)r^{2}+N^{2}\right) ,  \nonumber \\
\Omega _{4} &=&(r^{2}-N^{2})\left\{ \Lambda \frac{(r^{2}-N^{2})^{2}}{4k(k-1)}%
-\frac{(r^{2}-N^{2})}{4(k-1)}-\frac{k\alpha }{2(k+1)}\right\}   \label{GamCP}
\end{eqnarray}
The solutions of Eq. (\ref{Eqd}) describe NUT solutions, if (i) $F(r=N)=0$,
(ii) $F^{\prime }(r=N)=[(k+1)N]^{-1}$, (iii) $h(r=N)=0$ and (iv) the charge $%
q$ is less than a critical value $q_{{\rm crit}}$, where $q_{{\rm crit}}$ is
the solution of the system of two equations (\ref{gnut}). The first
condition comes from the fact that all the extra dimensions should collapse
to zero at the fixed point set of $\partial /\partial \tau $, the second one
ensures that there is no conical singularity with a smoothly closed fiber at
$r=N$, the third one comes from the regularity of vector potential at \ $r=N$
and the fourth condition comes up since $r=N$ should be the event horizon.

Using these conditions, one finds that the solutions of the differential
equation (\ref{Eqd}) with $\Omega _{i}$'s of Eqs. (\ref{GamCP}) yield a
non-extreme NUT solution for any given (even) dimension $k\geq 2$ provided
\begin{equation}
V\equiv V_{n}=\frac{(-1)^{k}}{2\sqrt{\pi }N^{2k-1}}\frac{\Gamma (k-\frac{1}{2%
})}{\Gamma (k+1)}q,
\end{equation}
the mass parameter $m$ is fixed to be
\begin{eqnarray}
m_{n} &=&-\frac{(k-2)!2^{k-1}N^{2k-3}}{(k+1)(2k+1)!!}\{ 8(k+1)^{2}\Lambda
N^{4}+4k(k+1)(2k+1)N^{2}  \nonumber \\
&&-4k(2k+1)(2k-1)(k-1)\alpha \}-\frac{4\Gamma (k+\frac{1}{2})}{\sqrt{\pi }%
k\Gamma (k+1)}N^{1-2k}q^{2}  \label{mncp}
\end{eqnarray}
and $q<q_{{\rm crit}}$. This solution has no curvature singularity at $r=N $.

Solutions of Eqs. (\ref{Eqd}) and (\ref{GamCP}) for $m\neq m_{n}$ in any
dimension can be regarded as bolt solutions. The value of the bolt radius $%
r_{b}>N$ may be found from the regularity conditions (i) $F(r=r_{b})=0$ and $%
F^{\prime }(r_{b})=[(k+1)N]^{-1}$. Applying these for ${\cal B}={\Bbb CP}^{k}
$ gives the following equation for $r_{b}$
\begin{eqnarray*}
&&4(k+1)\Lambda N{r_{b}}^{4}+4k {r_{b}}^{3}-4(k+1)N\left[ k+2\Lambda N^{2}%
\right] {r_{b}}^{2}+4k\left[ 4(k-1)\alpha -N^{2}\right] r_{b}  \nonumber \\
&&+N\left[ 4(k+1)\Lambda N^{4}+4k(k+1)N^{2}-8(k-1)k^{2}\alpha \right]
-4(k+1)(2k-1)^{2}N(\frac{V_{b}}{r_{b}})^{2}(r_{b}^{2}-N^{2})^{2}=0
\end{eqnarray*}
where $V_{b}$ is the solution of $h(r_{b})=0$.

Next we consider the solutions with the base space ${\cal B}=T^{2}\times
...\times T^{2}$. The field equation is given by (\ref{Eqd}), where now
\begin{eqnarray}
\Omega _{1} &=&-\alpha \{(2k-1)r^{2}+3N^{2}\}F+\frac{(r^{2}-N^{2})^{2}}{%
4(k-1)},  \nonumber \\
\Omega _{2} &=&-\frac{\alpha }{2(r^{2}-N^{2})}\left\{
(2k-1)(2k-3)r^{4}+2(2k-7)N^{2}r^{2}+3N^{4}\right\} ,  \nonumber \\
\Omega _{3} &=&-\frac{(r^{2}-N^{2})[(2k-1)r^{2}+N^{2}]}{4(k-1)},  \nonumber
\\
\Omega _{4} &=&\Lambda \frac{(r^{2}-N^{2})^{3}}{4k(k-1)}  \label{GamTT}
\end{eqnarray}
The solutions of Eqs. (\ref{Eqd}) and (\ref{GamTT}) yield an extreme NUT
solution for any given even dimension provided $q<q_{{\rm crit}}$ and the
mass parameter $m$ is fixed to be
\begin{equation}
m_{n}=-\frac{(k+1)(k-2)!2^{k+2}}{(2k+1)!!}\Lambda N^{2k+1}-\frac{4\Gamma (k+%
\frac{1}{2})}{\sqrt{\pi }k\Gamma (k+1)}N^{1-2k}q^{2}  \label{mnTT}
\end{equation}
where in this case the spacetime has no curvature singularity at $r=N$. Also
one may find that the Euclidean regularity at the bolt requires the period
of $\tau $ to be
\begin{equation}
\beta =-\frac{4k\pi r_{b}^{3}}{({r_{b}^{2}}-N^{2})(\Lambda {r_{b}^{2}-(2k-1)}%
^{2}V_{b}^{2})}  \label{betTT}
\end{equation}
and can have any value from zero to infinity as $r_{b}$ varies from $N$ to
infinity, and therefore one can have bolt solution.

Finally, we consider the solution when ${\cal B}=S^{2}\times T^{2}\times
...\times T^{2}$. In this case the field has the same form as Eq. (\ref{Eqd}%
) with
\begin{eqnarray}
\Omega _{1} &=&-\alpha \{(2k-1)r^{2}+3N^{2}\}F+(r^{2}-N^{2})\{\frac{\alpha}{k%
} +\frac{r^{2}-N^{2}}{4(k-1)}\},  \nonumber \\
\Omega _{2} &=&-\frac{\alpha }{2(r^{2}-N^{2})}\left\{
(2k-1)(2k-3)r^{4}+2(2k-7)N^{2}r^{2}+3N^{4}\right\} ,  \nonumber \\
\Omega _{3} &=&-\frac{(r^{2}-N^{2})[(2k-1)r^{2}+N^{2}]}{4(k-1)}+\frac{4(k-1)%
}{k}\alpha \{(2k-3)r^{2}+N^{2}\},  \nonumber \\
\Omega _{4} &=&\frac{(r^{2}-N^{2})^{2}[\Lambda (r^{2}-N^{2})-1]}{4k(k-1)}
\label{GamST}
\end{eqnarray}
Solutions of Eqs. (\ref{Eqd}) with (\ref{GamST}) yield a NUT solution for
any given even dimension with curvature singularity at $r=N$, provided the
mass parameter $m$ is fixed to be
\begin{equation}
m_{n}=-\frac{(k-2)!2^{k+1}}{(2k+1)!!}N^{2k-1}\left\{ 2(k+1)\Lambda
N^{2}+(2k+1)\right\} -\frac{4\Gamma (k+\frac{1}{2})}{\sqrt{\pi }k\Gamma (k+1)%
}N^{1-2k}q^{2}  \label{mnST}
\end{equation}
and $q<q_{{\rm crit}}$. Also one may find that the Euclidean regularity at
the bolt requires the period of $\tau $ to be
\begin{equation}
\beta =\frac{4k\pi r_{b}^{3}({r_{b}}^{2}-N^{2}+\frac{4(k-1)}{k}\alpha )}{{%
r_{b}^{2}}({r_{b}^{2}}-N^{2})[1-\Lambda ({r_{b}}^{2}-N^{2})]+{(2k-1)}%
^{2}V_{b}^{2}({r_{b}^{2}}-N^{2})^{2}}  \label{betST}
\end{equation}
Again, $\beta $ of Eq. (\ref{betST}) can have any value from zero to
infinity as $r_{b}$ varies from $N$ to infinity, and therefore one can have
bolt solution.

The asymptotic behavior of all of these solutions is locally AdS for\ $%
\Lambda <0$ provided $\left| \Lambda \right| <k(2k+1)/[(k-1)(2k-1)\alpha ]$
locally dS for $\Lambda >0$ and locally flat for $\Lambda =0$ . All the
different $F(r)$'s for differing base spaces have the same form as $\alpha $
goes to zero, while they reduce to the solutions of Gauss-Bonnet gravity
constructed in \cite{Deh1} when $q=V=0$.

\section{Concluding Remarks \label{con}}

We have presented a class of ($2k+2$)-dimensional Taub-NUT/bolt solutions in
Gauss-Bonnet-Maxwell gravity with cosmological term. These solutions are
constructed as $S^{1}$ fibrations over even dimensional spaces that in
general are products of Einstein-K\"{a}hler spaces. We found that the
function $F(r)$ of the metric depends on the specific form of the base
factors on which one constructs the circle fibration, while the form of
electromagnetic field is independent of the base space. This is different
from the solution of the Einstein-Maxwell gravity where the metric in any
dimension is independent of the specific form of the base factors. In the
presence of electromagnetic field, there exist two extra parameters, in
addition to the mass and the NUT charge, namely; the electric charge $q$ and
the potential at infinity $V$.

We found that in order to have NUT charged black holes in
Gauss-Bonnet-Maxwell gravity, in addition to the two conditions of uncharged
NUT solutions, there exists two other conditions. The first extra condition
comes from the regularity of vector potential at $r=N$ which gives a
relation between $q$ and $V$. Indeed, the existence of the parameter $V$
enables us to get a regularity condition on the one-form potential which is
identical to that required to obtain a NUT solution. If one of these
parameters vanishes then the other one should be equal to zero and the
solution reduces to the uncharged solution. The second extra condition comes
from the fact that the horizon at $r=N$ should be the outer horizon of the
black hole. Indeed, Gauss-Bonnet-Maxwell gravity admits NUT black holes
provided the charge parameter is less than a critical value $q_{{\rm crit}}$%
, which may be obtained by solving the system of two equations (\ref{gnut}).
In any dimension, the mass parameter $m$ which is fixed by these four NUT
conditions depends on the fundamental constant $\Lambda $, $\alpha $, $N$
and $q$.

We also found that when Gauss-Bonnet gravity admits non-extremal NUT
solutions with no curvature singularity at $r=N$, then there exists a
non-extremal NUT solution in Gauss-Bonnet-Maxwell gravity too. In $(2k+2)$%
-dimensional spacetime, this happens when the metric of the base space is
chosen to be ${\Bbb CP}^{k}$. Indeed, Gauss-Bonnet-Maxwell gravity does not
admit non-extreme NUT solutions with any other base space. We confirm that
when the base space has at most a 2-dimensional curved factor space with
positive curvature, then Gauss-Bonnet-Maxwell gravity admits extremal NUT
solutions as in the case of uncharged solutions. Finally, we obtained the
bolt solutions of Gauss-Bonnet-Maxwell gravity in various dimensions and
different base spaces, and gave the equations which can be solved for the
horizon radius of the bolt solution.

Although, we obtained the explicit form of the solutions in $6$, $8$ and $10$
dimensions, one can generalize these solutions in a similar manner for even
dimensions higher than ten. We gave the vector potential and the
differential equation of the function $F(r)$ in $2k+2$ dimensions. In ($2k+2$%
)-dimensional spacetime, we have only one non-extremal NUT solution with $%
{\Bbb CP}^{k}$ as the base space, and two extremal NUT solutions with the
base spaces $T^{2}\times T^{2}\times .....\times T^{2}$ and $S^{2}\times
T^{2}\times T^{2}\times .....\times T^{2}$. There is no curvature
singularity for the first two case, while for the latter case, the spacetime
has curvature singularity at $r=N$.

Insofar we see that the corrections of low energy limit of string theory
single out a preferred base space in order to have NUT solutions. Thus, the
investigation of the existence of NUT solutions in dimensionally continued
gravity, or Lovelock gravity with higher order terms might provide us with a
window on some interesting new corners of higher order gravity. Also, the
study of thermodynamic properties of these solutions remains to be carried
out in the future.

\begin{acknowledgements}
This work has been supported by Research Institute for Astrophysics and
Astronomy of Maragha, Iran.
\end{acknowledgements}

\end{document}